\documentclass[11pt,twoside]{article}
\usepackage{asp2014}

\aspSuppressVolSlug
\resetcounters

\begin{document}

\title{A scalable transient detection pipeline for the Australian SKA Pathfinder VAST survey}

\author{Sergio~Pintaldi$^1$, Adam~Stewart$^2$, Andrew~O'Brien$^3$, David~Kaplan$^3$ and Tara Murphy$^{2,4}$}
\affil{$^1$Sydney Informatics Hub, The University of Sydney, NSW 2006, Australia; \email{sergio.pintaldi@sydney.edu.au}}
\affil{$^2$Sydney Institute for Astronomy, School of Physics, The University of Sydney, NSW 2006, Australia}
\affil{$^3$University of Wisconsin-Milwaukee, Department of Physics, Milwaukee, WI, USA}
\affil{$^4$ARC Centre of Excellence for Gravitational Wave Discovery (OzGrav), Hawthorn, Victoria, Australia}

\paperauthor{Sergio~Pintaldi}{sergio.pintaldi@sydney.edu.au}{0000-0003-3860-5825}{University of Sydney}{Sydney Informatics Hub}{Sydney}{NSW}{2000}{Australia}
\paperauthor{Adam~Stewart}{adam.stewart@sydney.edu.au}{0000-0001-8026-5903}{University of Sydney}{Sydney Institute for Astrophysics, School of Physics}{Sydney}{NSW}{2000}{Australia}
\paperauthor{Andrew~O'Brien}{obrienan@uwm.edu}{0000-0003-4609-2791}{University of Wisconsin-Milwaukee}{Department of Physics}{Milwaukee}{WI}{53201}{USA}
\paperauthor{David~L.~Kaplan}{kaplan@uwm.edu}{0000-0001-6295-2881}{University of Wisconsin-Milwaukee}{Department of Physics}{Milwaukee}{WI}{53201}{USA}
\paperauthor{Tara Murphy}{tara.murphy@sydney.edu.au}{0000-0002-2686-438X}{University of Sydney}{Sydney Institute for Astrophysics, School of Physics}{Sydney}{NSW}{2000}{Australia}

\begin{abstract}
The Australian Square Kilometre Array Pathfinder (ASKAP) collects images of the sky at radio wavelengths with an unprecedented field of view, combined with a high angular resolution and sub-millijansky sensitivities. The large quantity of data produced is used by the ASKAP Variables and Slow Transients (VAST) survey science project to study the dynamic radio sky. Efficient pipelines are vital in such research, where searches often form a `needle in a haystack' type of problem to solve. However, the existing pipelines developed among the radio-transient community are not suitable for the scale of ASKAP datasets.

In this paper we provide a technical overview of the new \emph{``VAST Pipeline''}: a modern and scalable Python-based data pipeline for transient searches, using up-to-date dependencies and methods. The pipeline allows source association to be performed at scale using the {\tt Pandas} {\tt DataFrame} interface and the well-known {\tt Astropy} crossmatch functions. The {\tt Dask} Python framework is used to parallelise operations as well as scale them both vertically and horizontally, by means of a cluster of workers.
A modern web interface for data exploration and querying has also been developed using the latest {\tt Django} web framework combined with {\tt Bootstrap}.
\end{abstract}

\section{Introduction}
The ASKAP Survey for Variables and Slow Transients \citep[VAST;][]{item_vast} is the study of astrophysical transient and variable phenomena at radio wavelengths, such as flare stars and supernovae, using the new Australian Square Kilometre Array Pathfinder (ASKAP; Hotan et al. \emph{submitted}) telescope. The large field of view of ASKAP means that large areas of the radio sky can be surveyed regularly at sub-millijansky sensitivities. This has not been possible with previous radio telescopes, and means that ASKAP is now providing an unprecedented view of the dynamic radio sky. For example, the first shallow all-sky survey completed by ASKAP --- the Rapid ASKAP Continuum Survey \citep[RACS;][]{item_racs} --- provided 2.8 million source measurements, compared to 0.2~million sources detected in the previous best survey of the southern sky \citep[SUMSS;][]{item_sumss}.  This creates a data challenge for VAST, where regular epochs of the sky will result in the need to construct lightcurves for millions of astrophysical sources, while also being able to swiftly identify the small percentage of sources that exhibit transient behaviour, in addition to providing a visualisation solution for such a large and rich dataset.

Searching for transient and variable sources in image-plane radio data has unique challenges compared to other wavelengths, such as optical astronomy, because of inconsistencies that can occur between images. This means that methods such as `image differencing' can be inaccurate or difficult to produce, especially for short integration observations. Hence, a common technique is to perform source extraction on the images and to then associate the extractions into unique source time series data which can then be analysed for variability and transient behaviour. Currently, the main open-source software for image-based transient detection well-known to the radio astronomy community is the LOFAR Transients Pipeline \citep[\emph{``TraP''};][]{item_trap}. The TraP is a robust and powerful software package that can be used generally with radio image data from any radio facility, as demonstrated by previous successful transient and variable searches with data from a range of telescopes \citep{item_lofar_trap, item_meerkat_trap, item_chiles_trap}. However, when using the TraP with ASKAP survey data such as the VAST pilot survey, it became apparent that due to the scale of the data, the TraP performance did not meet our requirements. For example, a five image, full sensitivity, ASKAP dataset, i.e. five images of the same 30~deg$^2$ region of sky each with approximately 30\,000 sources, has a processing time of 95~min. The main issues were:
\begin{itemize}
\item TraP uses the PySE \citep{item_pyse} source extractor to measure the properties of the sources found in the images. However, source catalogues are a standard data product of the ASKAP processing pipeline, hence this step was no longer required and could potentially save a significant amount of processing time, although, the ability to read in catalogues is not a feature in TraP (v4.0).
\item TraP also has a `forced extractions' feature, this is when PySE is used to forcefully extract a Gaussian measurement of a source that was not detected in a certain image in order to `fill in' the gap in the light curve. The high source count with ASKAP meant that this step could also take considerable processing time.
\item Most of source association operations are performed in the database, using the {\tt SQL} language, which can become slow when there are high number of sources. In addition the logic written in {\tt SQL} makes debugging difficult.
\item The software is written in {\tt Python 2.7}, which reached end of life in January 2020, meaning that improvements gained by switching to Python 3 were not easily reachable without a considerable migration effort.
\end{itemize}

To solve these issues, we have developed a modern and scalable Python code base that incorporates and builds upon TraP features to efficiently process ASKAP data. The modern technology stack allows us to perform fast mass source association using {\tt Pandas} {\tt DataFrame} and well-known {\tt Astropy} crossmatch functions. We have also developed a web interface under the same code code base, from which a user can run the pipeline and importantly, visualise and query the results.

The main features of the developed pipeline are discussed in Section 2, while an overview of its architecture is illustrated in Section 3.

\section{Pipeline Processing and Features}
\label{pipeline_features}
The main objectives of the pipeline is to take the ASKAP image and associated data products (refer to Section~\ref{sec:inputs}) and perform the following the steps:
\begin{itemize}
    \item load and ingest the ASKAP images and related files;
    \item associate all the source measurements into single astronomical objects;
    \item perform force extractions to fill in non-detected sources;
    \item calculate aggregate metrics for the objects;
    \item output the results to a database and parquet files.
\end{itemize}

The main \emph{VAST Pipeline} technology stack is as follows: {\tt Python 3.6+}, {\tt Postgres 12+}, {\tt Astropy 4+}, {\tt Django 3+}, {\tt Dask 2+} and {\tt Bootstrap 4}. The following sections describe each step of the pipeline in more in detail along with other pipeline components.

\subsection{Image Ingestion and Inputs}
\label{sec:inputs}
Raw data from the telescope is processed by the ASKAP observatory and we require four of the standard data products: (i) the image file (FITS), (ii) the source catalogue with the extracted source measurements from the image (text), (iii) the noise file (FITS) and (iv) the background map file (FITS). Data products ii, iii and iv and produced by ``Selavy'' \citep{item_selavy}, which is the default source-finder software for the ASKAP images. In the ingestion phase, the pipeline reads the image file FITS header for meta data information such as the date of the observation and sky location, performs some data harmonisation operations on the source catalogue (e.g. error correction), and uploads the measurements to the database and also writes the measurements to parquet files, and finally uses the noise image to gain background root-mean-squared statistics of the radio image. Once an image has been ingested by the pipeline the information is always retained and the image is available for any other pipeline runs, i.e. measurements are never duplicated in the database.

\subsection{Source Association}
The source association operations make use of {\tt Astropy} \citep{item_astropy2013} cross-match and sky search functions, while all the data manipulations are performed in {\tt Pandas} {\tt DataFrame} \citep{item_pandas}. In general, the association process follows the same logic as that used by the TraP. This includes using the weighted average right ascension and declination values when associating and the handling of `one-to-many', `many-to-one' and `many-to-many' association types (a standard association is defined as `one-to-one'). We refer the reader to Section~4.4 of \citep{item_trap} for a detailed explanation of the association steps and types. We have implemented a minor improvement in that any relationships formed between sources because of the association types are recorded as part of the pipeline output. There are currently three association methods available:
\begin{enumerate}
    \item Basic: this mode uses the {\tt Astropy} crossmatch functions\\(\verb"match_to_catalog_sky") for a given fixed radius and provides only the closest match to each object. `One-to-many' association types are possible and are recorded.
    \item Advanced: it uses {\tt Astropy} search around the sky function (\verb"search_around_sky") in which all matching sources are found within the user specified search radius, i.e. not only the closest as in the basic method. `One-to-many', `many-to-one' and `many-to-many' association types are possible and are recorded. `Many-to-many` associations are reduced to either `one-to-one' or `one-to-many' association types.
    \item de~Ruiter: in this mode the association is performed similarly to the advanced mode, but the de~Ruiter radius \citep{item_scheers} is calculated and used in the source association process using a user specified de~Ruiter radius limit. The same association types as found in advanced are also possible in this mode.
\end{enumerate}
By default the association is performed on an image by image basis in observation date order, however two features exist to help decrease the processing time in some cases. First, it is possible to run association in parallel if the data set has images from distinct separate regions of the sky to analyse. For example, if a survey consisted of a sequence of images over time centred at RA=0 and Dec=0 and another sequence of images at RA=0 and Dec=--60, the pipeline can associate these two regions separately (in parallel) and collect the results at the end.

The second mode available is the ability to group images together to form a single epoch. Surveys such as the VAST Pilot Survey commonly mosaic a large region of the sky in one observing session, and then will observe the same area of sky again at a later time in another session. Depending on the science case, it can  beneficial to consider each observing session as one complete epoch to associate to the following epoch obtained from the next session - this is instead of considering each individual image that makes up the full survey area, i.e. the default method. Users are able to define which images should be considered as a single epoch for which the pipeline will then merge the source catalogues and remove duplicate measurements for each epoch and perform association on the combined catalogues. For example, the VAST Pilot survey phase 1 consists of 12 epochs, split over 6 distinct sky regions, and contains 8.3~million individual source measurements. If parallel association is used with the default image-by-image association then the total association step processing time is 150~min. If the epoch mode is activated grouping together the images from each of the 12 epochs, the association step time is reduced to 11~min.

\subsubsection{Ideal Coverage and New Source Analysis}
\label{sec:new_source}
Following association the pipeline calculates the ideal coverage each source should have, which allows the pipeline to determine: (i) which sources have a non-detection in an image where the location of the source is covered, and (ii) which sources are flagged as `new' sources. A `new' source is a source that was not present in the first observation of the data set and has a signal-to-noise ratio such that it should have been detected. These are important to flag as they are likely transient sources.

Sources in group (i) that have non-detections are passed onto the forced extraction step (see Section~\ref{sec:forced_extractions}). For new sources --- group (ii) --- a few more steps are taken to provide useful metrics for post-analysis. In the initial determination step above, the source is flagged as new if the signal-to-noise ratio (SNR) of the first measurement's peak flux in time is above a user defined threshold (default of 5) compared to the lowest RMS value recorded in the image that provided the non-detection. However, the RMS values of different regions in the image can vary drastically, hence, to help the user decide if the `new' source is significant, for each new source the pipeline measures the RMS value of the non-detection image at the precise location of the source in question. The SNR is then recalculated using the precise RMS value and is attached to the source in the pipeline outputs. This helps the end-user filter out weak `new' sources that are likely to be erroneous.

\subsection{Forced Extraction Measurements}
\label{sec:forced_extractions}
The sources that have a non-detection in an image where they should have been detected, group (i) sources in Section~\ref{sec:new_source}, are submitted to have measurements extracted from the image in question. The pipeline has an option to place a user-defined minimum threshold (default of 3) of the SNR of the source for the forced extraction to take place. The SNR determined using the peak flux of the first measurement in time against the minimum RMS value of the non-detection image. This provides some control over disregarding forced extractions for sources that are not expected to have been detected even in the best region of the image. This helps lower the computation time, especially for data sets with images with vastly different median RMS values.

The forced extractions are performed using a custom written function that uses a Gaussian the same size as the point-spread-function of the telescope in the observation. Extractions from images are also performed in parallel using {\tt Dask}.

\subsection{Aggregate and Variability Metric Calculations}
The pipeline calculates various aggregate statistics for each unique source, such as, for example, maximum and minimum flux values, total number of true detection and forced measurements, the average compactness (flux$_{int}$/flux$_{peak}$ and the number of relations. Two variability metrics are also included in the calculations - the $V$ (coefficient of variation) and $\eta$ (weighted reduced $\chi^2$ significance of the variation) metrics, as described in \citep{item_trap}. These are calculated for both the integrated and peak flux values.

In addition to the $\eta$ and $V$ metrics the pipeline also calculates the metrics commonly used to perform a `two-epoch' variability analysis \citep{item_two_epoch_analysis}, which has the advantage of being able to define variability on more specific timescales. For each source, the $V_s$ and $m$ metrics are calculated for every unique pair of measurements at different timestamps. The metrics for each unique pair in the data set are uploaded the database and written to parquet. The pipeline also uses a user-defined threshold $V_s$ value (default of 4.3) to assign the most significant metric pair to the source itself to allow for swift filtering on sources themselves. For example if a source has pairs of $V_s = 4.3$, $m = 0.8$, and $V_s = 4.3$, $m = 1.2$ and $V_s = 3.0$, $m = 0.5$, the pipeline assigns the $V_s = 4.3$, $m = 1.2$ pair to the source.

\subsection{Computation Cluster}
The {\tt Dask} framework is used to parallelise operations on {\tt DataFrames} such as the computation of metrics, merging {\tt DataFrames}, and groupby-apply operations. It is also used by the parallel source association mode, and in the forced source extractions, among other minor pipeline features. The use of {\tt Dask} enables out-of-RAM operations as {\tt Dask} is able to manage the RAM limits, and more importantly it provides horizontal scalability with the use of a cluster of workers.

\subsection{Web App}
The results of a pipeline run can be explored in a website developed using the {\tt Python Django} framework, in which a {\tt Bootstrap4} template was used. A {\tt Postgres12+} database is used to serve the data to the web app, and to store the data model as well as relationships between the data entities (e.g. sources and images). The {\tt Q3C} {\tt Postgres} plugin \citep{item_q3c} is used to add cone search functionality in the data query. The interface enables:
\begin{itemize}
    \item Displaying general statistics about all the pipeline runs (e.g. total amount of images processed);
    \item Inspecting and configuring single pipeline runs: e.g. change the configuration file, initialise a pipeline run with chosen parameters and list of images to process;
    \item Start or re-run a pipeline run with the computation in the background. This functionality is enabled by {\tt Django} extension app {\tt Django-Q};
    \item Displaying tables related to images, measurements, sources, pipeline runs;
    \item Displaying detailed image information: name, positions, axis and RMS; sky region preview with {\tt Aladin Lite} \citep{item_aladin}, user comments; extracted measurements table;
    \item Displaying detailed measurement information: position data, axis and other data; sky region preview with {\tt Aladin Lite}; postage stamp cutout using {\tt JS9} \citep{item_js9}; user comments; sources and siblings tables;
    \item Displaying detailed source information: source metrics and data such as position, flux, $\eta$ and \emph{V} metrics, light curve using {\tt Bokeh} \citep{item_bokeh}, external catalogue crossmatches and references using {\tt Astroquery} package \citep{item_astroquery}, postage stamp cutouts of the associated measurements (using the {\tt JS9} library), related sources, measurements (the single sources data points over time associated into the same astronomical object), sky region preview (using {\tt Aladin Lite} service);
    \item Exporting tables to excel or CSV data format;
    \item Commenting and tagging: for example a source can be tagged as a particular astronomical object;
    \item ``Starring'' a source: a user can add a source to their own favourite sources list, with a comment;
\end{itemize}

\subsection{Parquet Outputs}
All outputs of a pipeline run are also saved in parquet files for raw data access and computation in the {\tt Dask} cluster. Saving in parquet format also makes the results be easily portable and accessible, in that they can easily be transported to local users own machines and, for example, be loaded into a Jupyter Notebook environment using packages such as {\tt Pandas} or {\tt Vaex} \citep{item_vaex} for offline post-processing.

\articlefigure{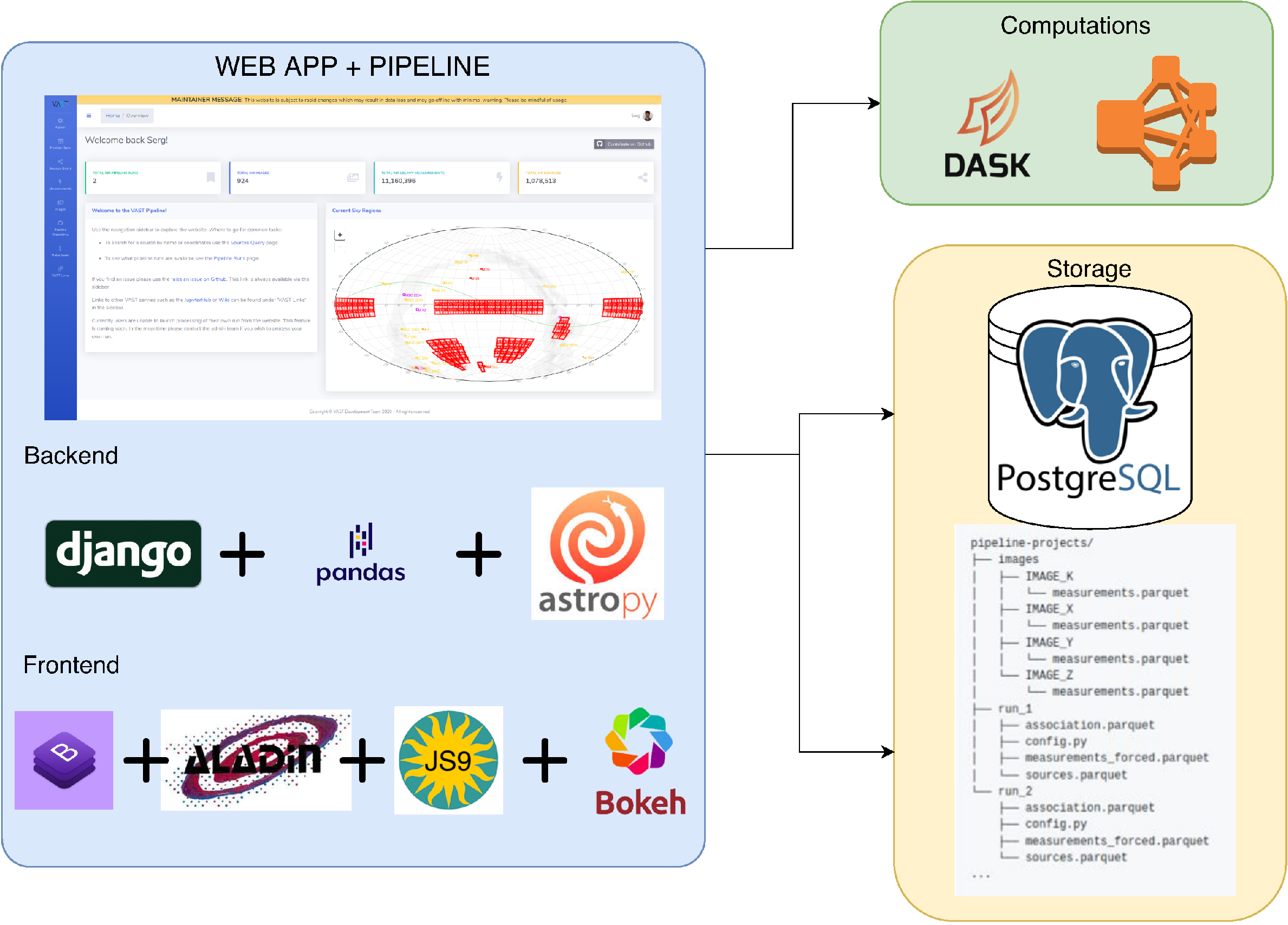}{arch_fig1}{VAST Pipeline architecture and stack.}

\section{Pipeline Architecture}
\label{pipeline_architecture}
A general schematic of the pipeline architecture is shown in figure \ref{arch_fig1}. The back-end as well as the front-end stacks are shown, on top of the storage and the computing platforms. The pipeline is a web application developed in Python using the Django web framework. Although {\tt Apache Hadoop} and its ecosystem of tools \citep{item_review_hadoop}, such as {\tt Apache Spark}, are the most predominant tools in ``Big Data'' applications, we chose to adopt Python and its eco-system of tools (i.e. {\tt Pandas} {\tt DataFrame} and {\tt Dask}) as the main pipeline technology stack due to their simplicity to develop and maintain, and popularity among the scientific community, in addition to astronomy specific libraries such as {\tt Astropy}. {\tt Dask} provides similar features of Spark, both from a data processing as well as a scalability point of view.\footnote{Dask vs Spark: \url{https://docs.dask.org/en/latest/spark.html}}

The basic data relationship between images and sources is represented in figure \ref{relation_fig2}:
\begin{itemize}
    \item Image: the ASKAP sky image taken at a specific time, with the main image file and related noise, background map and extracted measurements files.
    \item Measurement: a single source measurement at a specific time. Multiple measurements belong to a single image.
    \item Source: collection of associated measurements over time and images, that belong to the very same unique astronomical object.
\end{itemize}
\articlefigure{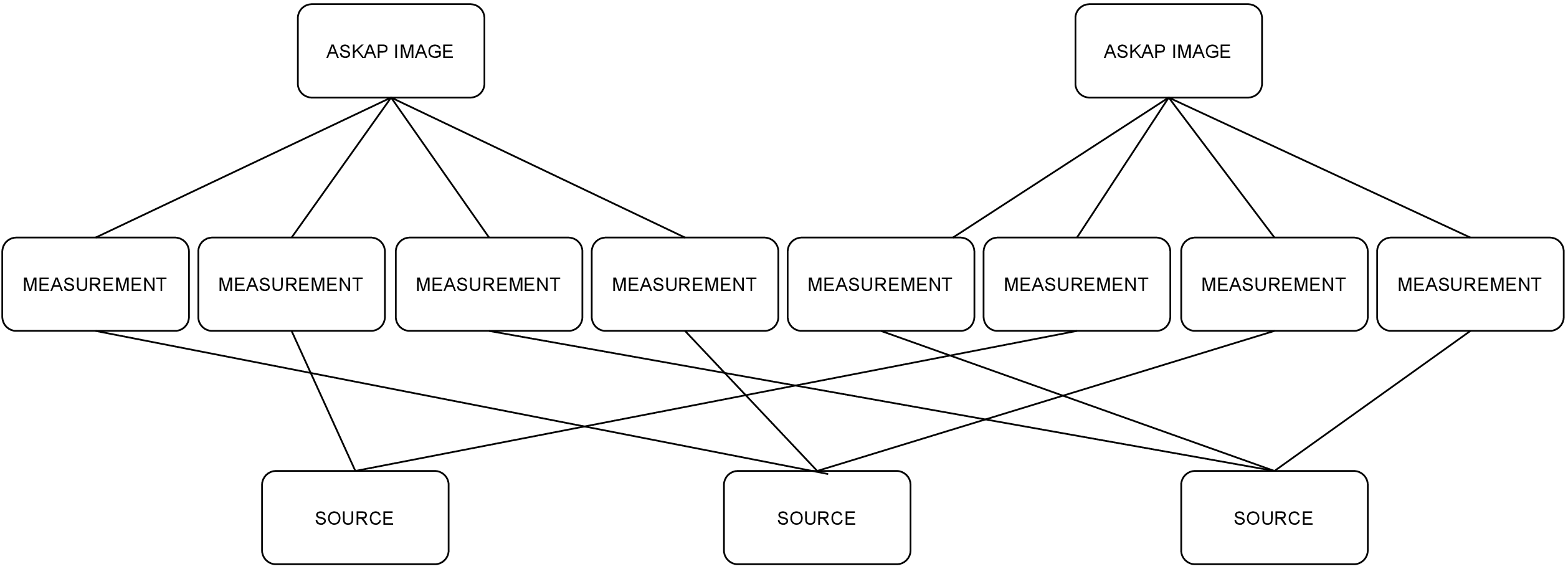}{relation_fig2}{VAST Pipeline image data relationship.}
This data schema avoids duplication of data when processing the images with difference settings. For example, when using two different choices for the association radius, the pipeline uses the same measurements objects, but will create different source objects due to the different association settings. This design choice was possible as the source finder is external to the pipeline. On the other hand, such design still enables processing an image with two different source-finder settings. In fact they will treated as two different image entities in the pipeline, though they are the same physical image.

\section{Summary}

The VAST Pipeline is a modern and scalable detection software for the discovery of radio transient and variable sources in data from the ASKAP telescope. It is built upon legacy community developed software, but with a modern and simpler architecture. The results have been validated against the ones produced by ``Trap'', and by individual custom-developed code bases, developed by researchers and students at the University of Sydney. It is able to process the VAST Pilot Survey phase 1 data (924 images, 8.3~million measurements, and $\sim$1 million sources), using de~Ruiter association, in approximately 13 hours on a machine with 16 CPUs and 64 GB of RAM. The processing time includes uploading of data to database, the association step, aggregate and variability metrics calculation and analysis (two-epoch analysis, `new sources', related sources), as well as forced extraction.
The adoption of a modern technology stack will ensure a long life expectancy of the VAST Pipeline.

\section{Acknowledgements}
This tool was developed in collaboration with the Sydney Informatics Hub (SIH), a core research facility at the University of Sydney. DK and AO are supported by NSF grant AST-1816492. TM acknowledges the support of the Australian Research Council through grants FT150100099 and DP190100561.
Parts of this research were conducted by the Australian Research Council Centre of Excellence for Gravitational Wave Discovery (OzGrav), project number CE170100004. The authors thanks the creators of \emph{SB Admin 2} bootstrap dashboard to make the template freely available.

\clearpage

\bibliography{mybib}

\end{document}